\documentclass[a4paper,11pt]{article}
\usepackage[utf8]{inputenc}
\usepackage{lmodern}
\usepackage[T1]{fontenc}
\usepackage{amsmath,amsthm,amsfonts,amssymb,bm}
\usepackage{enumitem}
\usepackage{graphicx}
\usepackage{subcaption}  
\usepackage{booktabs}
\usepackage{xcolor}
\usepackage{authblk}
 \usepackage{setspace}
\usepackage[authoryear]{natbib}
\usepackage[right=2.5cm,left=2.5cm,top=2cm,bottom=3cm]{geometry}
\usepackage[font={small,it}]{caption} 
\usepackage[colorlinks,linkcolor=blue,citecolor=blue,urlcolor=blue]{hyperref}
\usepackage{mathtools}

\usepackage{multirow}
\usepackage{makecell} 
\usepackage{afterpage}
\usepackage{float}

\usepackage{color}

\newtheorem{theorem}{Theorem}

\title{ \textsc{Confidence set for mixture order selection} }
\author[1]{Alessandro Casa}
\author[1]{Davide Ferrari}
\affil[1]{Faculty of Economics and Management, Free University of Bozen-Bolzano}
\date{}                     

\begin{document}

\doublespacing

\maketitle

\begin{abstract}
A fundamental challenge in approximating an unknown density using finite Gaussian mixture models is selecting the  number of mixture components, also known as order. Traditional approaches choose a single best model using information criteria. However, often models with different orders yield similar fits, leading to substantial model selection uncertainty and making it challenging to identify the optimal number of components. In this paper, we introduce the Model Selection Confidence Set (MSCS) for order selection in Gaussian mixtures -- a set-valued estimator that, with a predefined confidence level, includes the true mixture order across repeated samples. Rather than selecting a single model, our MSCS identifies all plausible orders by determining whether each candidate model is at least as plausible as the best-selected one, using a screening based on a penalized likelihood ratio statistic. We provide theoretical guarantees for asymptotic coverage, and demonstrate its practical advantages through simulations and real data analysis.
\end{abstract}
\smallskip
\noindent \textbf{Keywords:} Finite mixture models, order selection, penalized likelihood ratio test, model selection confidence sets, model selection uncertainty.

\section{Introduction}\label{sec:introduction}

Finite Gaussian mixture models are a powerful tool for approximating complex probability distributions and capturing heterogeneity in random phenomena; see \citet{fruhwirth2019handbook,yao2024mixture} for comprehensive reviews. They are widely used across scientific fields for semiparametric density estimation and model-based clustering \citep[see, e.g.,][]{roeder1997practical,bouveyron2019model}. In both settings, selecting the correct mixture order is essential, as misestimation can cause identifiability issues and distort the assessment of heterogeneity, ultimately affecting cluster estimation and model interpretation. An under-specified model may fail to capture important data structures, while an over-specified one may introduce spurious modes or noise sensitivity.


Traditionally, order selection is carried out according to the single-best-model paradigm, in which a single optimal number of components is chosen based on a predefined selection criterion. Existing strategies fall into three main categories: hypothesis testing \citep{mclachlan1987bootstrapping}, penalized likelihood methods \citep{chen2009order,manole2021estimating, huang2022statistical}, and information criteria; see \citet{celeux2019model} for a review. Information criteria are unarguably the most widely used, with the Bayesian Information Criterion \citep[BIC,][]{schwarz1978estimating} being a common choice. This popularity is due to its model-selection consistency property in regular settings \citep{keribin2000consistent}, and  numerical studies highlighting its superior performance over other approaches \citep{steele2010performance}. However, in relatively small samples, selection uncertainty can be substantial, making it difficult to distinguish between mixture models of different orders. When multiple models fit the data similarly well, committing to one may obscure meaningful structural variations and lead to either overfitting or underfitting.

This paper addresses model selection uncertainty by introducing the Model Selection Confidence Set (MSCS) for order selection. Instead of relying on a single selected model, our approach acknowledges multiple plausible models, enhancing interpretability and offering deeper understanding of underlying phenomenon. The MSCS is a set-valued estimator that, with a given confidence level, contains the true mixture order under repeated sampling. Like confidence intervals for parameters, it accounts for multiple plausible explanations of the data rather than enforcing a single choice. Its size reflects selection uncertainty: in larger samples, the MSCS is small, converging toward a single best model in extreme cases; in smaller samples it expands, thus indicating greater uncertainty in model selection. As the sample size grows, the MSCS captures the true order with high probability, providing a principled framework for handling order selection uncertainty.

Our methodology builds on the confidence set framework first introduced by \citet{ferrari2015confidence} for variable selection in linear models, which employs a screening approach based on the F-test. \citet{zheng2019model} extended their approach to more general likelihood settings using the likelihood ratio statistics to screen the useful models. While the latter provides asymptotic coverage for a broad class of parametric models, its extension to mixture models is nontrivial. Violations of regularity conditions, particularly due to parameters lying on the boundary of the parameter space, invalidate standard likelihood ratio tests for nested models, necessitating an alternative approach. We address this by leveraging distributional results of \citet{vuong1989likelihood} to construct a rigorous confidence set for the estimated mixture order.

\section{Confidence set construction}\label{sec:Confid_Set}

Let $X_1, \dots, X_n$ be an i.i.d. sample from a univariate random variable $X$ with unknown density $h(x)$ on $\mathbb{R}$. We use Gaussian mixtures as flexible approximating models rather than assuming they represent the true underlying distribution. Specifically, we approximate \( h \) using a \( k \)-component Gaussian mixture density of the form:
\begin{equation}\label{eq:eqMixture}
    f(x; \theta_k) = \sum_{j = 1}^k \pi_j \, \phi(x; \mu_j, \sigma_j^2),
\end{equation}
where \( \phi(x; \mu_j, \sigma_j^2) \) denotes the univariate Gaussian density with mean \( \mu_j \in \mathbb{R} \) and variance \( \sigma_j^2 > 0 \). The coefficient \( \pi_j \in (0,1) \) is the mixing proportion of the \( j \)-th component, with \( \sum_{j=1}^k \pi_j = 1 \). The full parameter vector is denoted by \( \theta_k = (\pi_1, \mu_1, \sigma_1^2, \dots, \pi_k, \mu_k, \sigma_k^2  )^\top \in \Theta_k \subset \mathbb{R}^{p_k}\), where \( p_k = 3k - 1 \) denotes the number of free parameters.  Since the components are Gaussian, the mixture density in~\eqref{eq:eqMixture} is invariant under permutations of the component labels in $\theta_k$. Extending the framework to multivariate mixtures is possible in principle but requires careful handling of additional technical challenges.

 A central objective in mixture modeling is to select the number of components  $k$ that best approximates the unknown density $h(x)$. To this end, we define the Kullback-Leibler (KL) divergence between \( h \) and the class of \( k \)-component mixtures, minimized over the parameter space, $
\text{KL}(k) := \inf_{\theta_k \in \Theta_k} \mathbb{E} \left[ \log \left\{ {h(X)}/{f(X; \theta_k)} \right\} \right]$. Here, and throughout the paper, all expectations are taken with respect to the data-generating process with density $h$. The optimal mixture order is defined as $
k_0 :=  \min\left\{ k \in \mathcal{K}:  \text{KL}(k) = \min_{\ell \in \mathcal{K}} \text{KL}(\ell) \right\}$,
where \( \mathcal{K} = \{1, \dots, k_{\max}\} \), \( k_{\max} < \infty \), denotes the set of candidate model orders. Our goal is to quantify the uncertainty in selecting the number of components by constructing MSCS, starting from a reference estimate $\hat{k}$ of $k_0$, which represents the selected model using an arbitrary selection criterion. Specifically, given a  confidence level  $(1-\alpha)$, with $0<\alpha<1$, we aim to construct a set of orders $\hat \Gamma$ such that
\begin{equation}\label{eq:MSCS_prob} \lim_{n \rightarrow \infty} \mathbb{P}\left(k_0 \in \hat\Gamma \right) \ge 1-\alpha,
\end{equation}
meaning that, asymptotically, $k_0$ is included in $\hat \Gamma$ with large probability. The MSCS  serves as a diagnostic tool for assessing model specification. If a model falls outside $\hat \Gamma$, it is likely either too simplistic, omitting essential components, or overly complex, incorporating unnecessary ones. The MSCS cardinality, $\text{card}(\hat \Gamma)$, reflects the amount of selection uncertainty: when the sample size is large, and models with different orders distinguishable, $\text{card}(\hat \Gamma)$ is expected to be small. In the most extreme case, $\hat \Gamma$ converges to the set $\{ k_0\}$ containing the single optimal order, as $n$ increases. Conversely, with smaller sample sizes $\text{card}(\hat \Gamma)$ increases, signaling greater uncertainty in model selection.

Operationally, we start with a reference model with $\hat k$ components, which is included by default in the MSCS. We then compare it to models of varying orders in  $\mathcal{K}$. For each $k \in \mathcal{K}$, we test the null hypothesis  $H_0: \text{KL}(k) = \text{KL}(\hat k)$, i.e., that the model of order $k$ and the selected model with $\hat{k}$ components are equally close to the truth, against the alternative $H_1: \text{KL}(k) > \text{KL}(\hat k)$, stating that the model of order $k$ is further from the truth compared to the selected model.

To carry out the hypothesis test, we use a Vuong-type test statistic \citep{vuong1989likelihood}. Let $d_i = \log f(X_i; \hat\theta_{\hat{k}}) - \log f(X_i; \hat\theta_{k})$ be the log-likelihood difference for the $i$th observation based on fitted Gaussian mixtures $ f(\cdot; \hat\theta_{\hat{k}})$ and $ f(\cdot; \hat\theta_k)$, having respectively, $\hat{k}$ and $ k$ components, with $\hat\theta_{\hat{k}}$ and $\hat\theta_k$ denoting the corresponding maximum likelihood estimators. The log-likelihood ratio is then given by $
\Lambda(\hat{k}, k) := \sum_{i=1}^n d_i = \ell(\hat\theta_{\hat{k}}) - \ell(\hat\theta_k)$, where \( \ell(\cdot) = \sum_{i=1}^n \log f(X_i; \cdot)\) denotes the log-likelihood for the entire sample. Define the penalized Vuong statistic as
\begin{equation}\label{eq:testStat}
V(\hat{k}, k) := \frac{ \Lambda(\hat{k}, k) - \delta_n(\hat{k}, k) }{ \sqrt{n} \, \hat\omega(\hat{k}, k) },
\end{equation}
where $\hat\omega_n^2(\hat{k}, k) := \ n^{-1} \sum_{i=1}^n ( d_i - \bar{d})^2$, and $\delta_n(\hat{k}, k)$ is a penalty term accounting for model complexity. The statistic \( V(\hat{k}, k) \) quantifies how much the reference model with \( \hat{k} \) components improves the fit to the data compared to an alternative model with \( k \) components, while adjusting for the complexity via \( \delta_n(\hat{k}, k) \). A large positive value indicates that the selected model fits significantly better than the alternative; conversely, a large negative value suggests that the alternative model with \( k \) components is superior.

In our setting, the models \( f(\cdot; \hat\theta_{\hat{k}}) \) and \( f(\cdot; \hat\theta_k) \) are finite Gaussian mixtures with different numbers of components, which are not nested in the classical sense. Although a \( k \)-component mixture can be embedded within a \( \hat{k} > k \) component model by shrinking some weights to zero or merging components, these extreme parameter configurations lie on the boundary of the parameter space and should be avoided since they invalidate standard likelihood ratio asymptotics. The framework of \citet{vuong1989likelihood} remains applicable under such non-nestedness and model misspecification: under the null hypothesis that both models yield equal KL divergence from the true density \( h \), the statistic \( V(\hat{k}, k) \) converges in distribution to a standard normal; under the alternative, it diverges to \( \pm \infty \), enabling model discrimination.

Given a significance level $\alpha$, the Model Selection Confidence Set (MSCS) is defined as
\begin{equation}
\hat \Gamma = \left\{ k \in \mathcal{K} : V(\hat{k}, k) \leq z_\alpha \right\} \cup \{ \hat{k} \},
\end{equation}
where \( z_\alpha \) denotes the upper \( \alpha \)-quantile of the standard normal distribution. That is, the MSCS consists of all candidate orders \( k \) for which there is no statistically significant evidence that the reference model \( \hat{k} \) fits the data better, according to the Vuong-type test, along with the reference model itself, which is included by construction.

The penalty term $\delta_n(\hat{k}, k)$ in \eqref{eq:testStat} balances model complexity and goodness-of-fit in finite samples. It corrects for the increased flexibility of models with a larger number of components, thus preventing the inclusion of overly complex models.   While  $\delta_n$  is allowed to grow with $n$, it must not affect the asymptotic behavior of the statistic under the alternative hypothesis. To  this end, we impose $n^{-1/2} \delta_n(\hat{k},k) = o_p(1)$, similarly to \citet[Eq. 5.10]{vuong1989likelihood}. This condition holds for common information criterion penalties, such as those based on AIC and BIC differences, leading to $\delta_n(\hat{k},k) = p_{\hat{k}} - p_k$ and $\delta_n(\hat{k},k) = \log(n)(p_{\hat{k}} - p_k)/2$, respectively. Takeuchi Information Criterion \citep[TIC,][]{takeuchi1976distribution} can also be considered, providing a robust correction under model misspecification. This leads to $\delta_n(\hat{k},k) = \tilde{p}_{\hat{k}} - \tilde{p}_k$, where $\tilde{p}_k = \text{tr} \{ A(\hat{\theta}_{k})^{-1} B(\hat{\theta}_{k}) \}$, with $A(\theta_k)= - \mathbb{E}[ \partial^2 \log f(X; \theta_k) / \partial \theta^2_k]$ and $B(\theta_k)= \mbox{Var}\left[ \partial \log f(X; \theta_k) / \partial \theta_k \right]$ representing the Fisher information matrix and the score variance matrix, respectively.

\section{Asymptotic coverage of the MSCS}\label{sec:theory}

In this section, we study the large-sample behavior of our MSCS, focusing on its coverage probability. For each $ k \in \mathcal{K}$, we define the pseudo-true parameter vector $\theta_k^\ast \in \Theta_k$ as the maximizer of the expected log-likelihood function $
L_k(\theta_k) = \mathbb{E}[\log f(X; \theta_k)]$. Theorem \ref{thm:coverage} establishes the asymptotic validity of the proposed procedure, showing that the optimal mixture order \( k_0 \) is contained in the MSCS with probability at least \( 1 - \alpha \) in large samples.

\begin{theorem}\label{thm:coverage}
Assume: (C1) \( X_1, \dots, X_n \) are i.i.d. with strictly positive density \( h(x) > 0 \) and finite fourth moment. For each \( k \in \mathcal{K} \): (C2) \( \Theta_k \) is compact; (C3) the pseudo-true parameter \( \theta_k^\ast \) is unique and lies in the interior of \( \Theta_k \); (C4) the Fisher information matrix \( A(\theta_k^\ast) \) is positive definite. Moreover, (C5) the penalty satisfies \( n^{-1/2} \delta_n(k_1, k_2) = o_p(1) \) for all \( k_1, k_2 \in \mathcal{K} \). Then the MSCS \( \hat \Gamma \) satisfies $
\lim_{n \to \infty} \mathbb{P}(k_0 \in \hat{\Gamma}) \geq 1 - \alpha$.
\end{theorem}

\begin{proof}
The total probability of not including the correct order $k_0$ in the MSCS is
    \begin{eqnarray}\label{eq:ProofTh1}
        \mathbb{P}(k_0 \notin \hat{\Gamma})  =  \mathbb{P}(k_0 \notin \hat{\Gamma} \vert \hat{k} = k_0)\mathbb{P}(\hat{k} = k_0) +  \mathbb{P}(k_0 \notin \hat{\Gamma} \vert \hat{k} \neq k_0)\mathbb{P}(\hat{k} \neq k_0).
    \end{eqnarray}
    Since $\hat{k}$ is included in $\hat{\Gamma}$ by default, the first term in (\ref{eq:ProofTh1}) is zero. When $\hat k \neq k_0$,  since $0 = \text{KL}(k_0)\leq \text{KL}(k)$ for all $k \in \mathcal{K}$, we have the worst-case bound
\begin{align}
\mathbb{P}(k_0 \notin \hat \Gamma \vert \hat{k} \neq k_0 ) = \mathbb{P}\left( V(\hat{k}, k_0)  > z_{\alpha} \vert \hat{k} \neq k_0\right)
 \leq
\max_{k: \text{KL}(k) \leq \text{KL}(\hat{k})}  \mathbb{P}\left( V(\hat{k}, k) > z_{\alpha}  \vert \hat{k} \neq k_0 \right). \label{eq:ineq1}
\end{align}

Conditions A1--A6 of \citet{vuong1989likelihood} hold under our assumptions  C1--C5. A1 follows from C1. A2(a) is implied by the structure of Gaussian mixtures and A2(b) follows by C2. A3(a) is satisfied since \( \log f(x; \theta_k) \) is bounded above by \( c_1 = \log(1/\sqrt{2\pi \sigma_{min}}) \), for some $\sigma_{min}>0$ and below by \( -c_2(1 + x^2) \), $c_2>0$, so that \( |\log f(x; \theta_k)| \leq \max \{ c_1, c_2(1 + x^2) \} \), which is integrable by C1; A3(b)   corresponds to C3. A4(a) is satisfied by the assumption of Gaussian components; A4(b) holds because, under compactness in C2, the score and Hessian of \( \log f(x; \theta_k) \) are polynomially bounded in $ x $, with their elements satisfying, respectively,
\( \left| \partial \log f(x; \theta)/ \partial \partial_{\theta_{k,r}} \cdot \partial \log f(x; \theta)/ \partial \partial_{\theta_{k,s}} \right| \leq c_3 (1 + x^4) \) and
\( \left| \partial^2 \theta_{k,s} \log f(x; \theta_k)/ \partial {\theta^2_{k,r}} \right| \leq c_4 (1 + x^4) \), for some constants $c_3, c_4$, with $\theta_{k_,r}, \theta_{k_,s}$ being elements of $\theta_k$; these bounds are integrable by C1. A5(a) and A5(b) are ensured by C3 and C4, respectively.  A6 follows by noting that under compactness (C2) $|\log f(x; \theta_k)|^2 \leq \max \{ c^2_1, c_2(1 + x^4 + 2 x^2) \}$ (see proof of A3(a) above), which is integrable by C1.

 If \( \text{KL}(k) < \text{KL}(\hat{k}) \),  Theorem 5.1 of \citet{vuong1989likelihood} implies \( V(\hat{k}, k) \to - \infty \) almost surely, so $\mathbb{P} \left( V(\hat{k}, k)\ > z_{\alpha}  \vert \hat{k} \neq k_0 \right) \to 0$, as $n \to \infty$. Thus, the maximum in \eqref{eq:ineq1} is asymptotically attained at \( k \) with \( \text{KL}(k) = \text{KL}(\hat{k}) \). Under the null $H_0: \mbox{KL}(k_1) = \mbox{KL}(k_2)$,   Theorem 5.1 of \citet{vuong1989likelihood} implies \( V(\hat{k}, k) \overset{d}{\to} \mathcal{N}(0,1) \). Hence, taking the limit in  (\ref{eq:ProofTh1}) yields
\begin{align}
\lim_{n \to \infty}   \mathbb{P}(k_0 \notin \hat{\Gamma}) & \leq \alpha \lim_{n \to \infty} \mathbb{P} \left(  \hat{k} \neq k_0 \right) \leq \alpha.
\end{align}
Rearranging gives  $
 \lim_{n \to \infty} \mathbb{P}(k_0 \in \hat{\Gamma}) \geq 1 - \alpha$ and completes the proof.
\end{proof}

We remark that Theorem 1 holds even when the initial reference order \( \hat{k} \) is misspecified. This robustness distinguishes MSCS from traditional single-model selection approaches that rely on model consistency or correct specification. While coverage is preserved by construction, the size of the MSCS reflects the power of the Vuong-type test and depends on the quality of the reference model \( \hat{k} \). When $ \hat{k}$ is close to $ k_0$, the MSCS is narrower and more informative. Conversely, if \( \hat{k} \) is far from \( k_0 \), the test may lack power to exclude incorrect alternatives, yielding a wider MSCS. Nonetheless, the coverage guarantee remains intact, underscoring the method's capacity to adapt to uncertainty without sacrificing inferential validity. Numerical results in Section~\ref{sec:Simulations} support this behavior.

Assumptions C1-C4 are standard in the asymptotic theory of likelihood-based estimation and model selection. Assumption C1 ensures well-defined log-likelihood contributions, while compactness of the parameter space (C2) guarantees the existence of maximum likelihood estimators and avoids pathological configurations. While compactness may appear restrictive---since parameter spaces in Gaussian mixture models are typically unbounded---it is a standard technical device that facilitates uniform convergence arguments and rules out degenerate configurations such as vanishing variances; e.g., see \citet{chen1995consistent}, \citet{nguyen2013convergence}. A key assumption is the uniqueness of the pseudo-true parameter \( \theta_k^\ast \) (C3), which ensures that the model of order \( k \) approximates the true density \( h(x) \) uniquely in terms of Kullback--Leibler divergence. In the case where \( h(x) \) is a finite Gaussian mixture, this condition typically holds under  first-order identifiability, as discussed in  \citet{ho2016strong}, provided the components are sufficiently distinct in both location and scale. Stronger identifiability conditions such as second-order identifiability, involving independence of score derivatives, are generally not satisfied by Gaussian mixtures, but this  does not affect the theoretical behavior of Vuong-type tests. Particularly, to ensure the null distribution of Vuong's statistic, our framework still requires a regular behavior of the Hessian of the log-likelihood locally at $\theta_k^\ast$ (C4) and boundedness of the log-likelihood differences, implied by existence of fourth moments in Condition C1.

\section{Numerical examples}\label{sec:Simulations}

\paragraph{Synthetic data.} To empirically evaluate the finite-sample performance of our approach, we generate $n$ i.i.d. samples from univariate Gaussian mixture densities used in \citet{Casa2020EJS}. The corresponding density functions and the associated parameter settings are summarized in Table~\ref{tab:tabSimulationSetting}. These densities pose challenges in density estimation, reflecting different modality patterns and interactions between modes and components. We consider sample sizes $ n = 100, 250, 1000$ and $\alpha = 0.01, 0.05, 0.1$. Results are based on the correction $\delta_n(\hat{k}, k) = p_{k}  -  p_{\hat k}$, with  reference order $\hat{k}$ selected via BIC.
  
\begin{table}[t]
\centering
\begin{minipage}{\textwidth}
  \centering
  \includegraphics[scale=0.63]{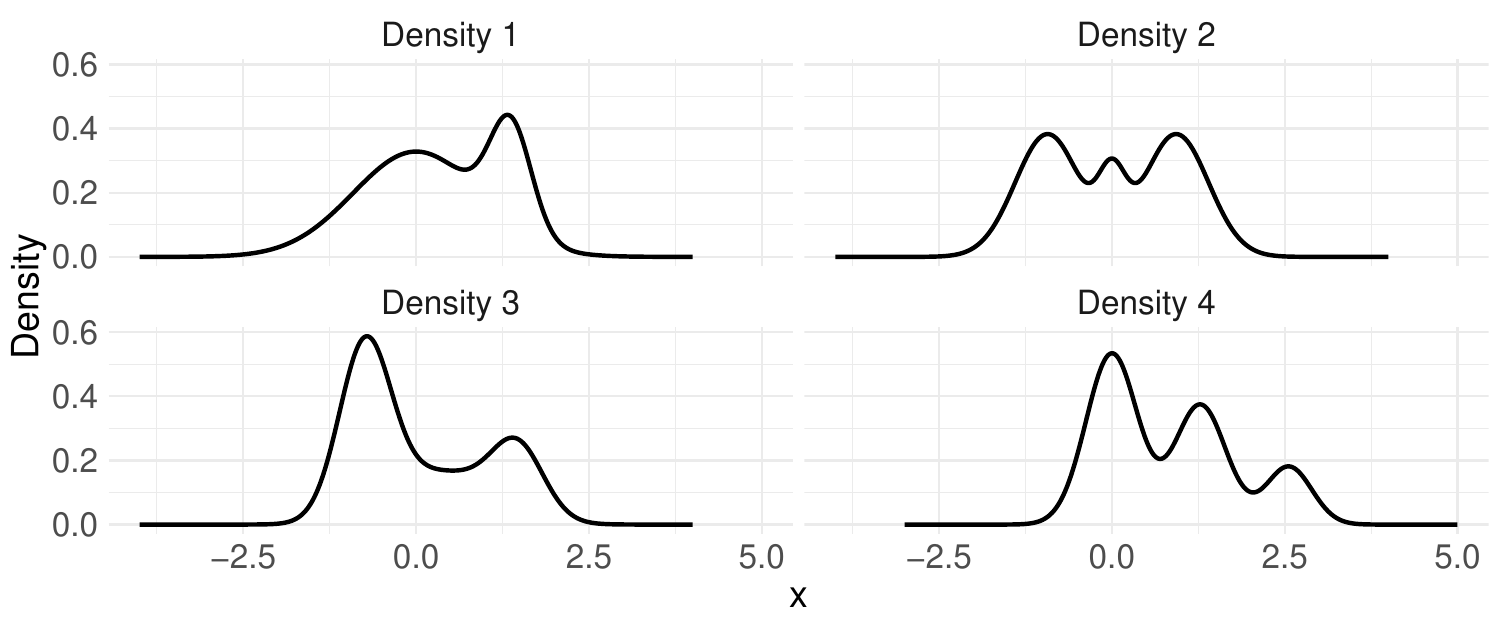}
  \vspace{0.3em}

   \begin{tabular}{c ccc c ccc c ccc}
    \hline
    & \multicolumn{3}{c}{Component 1} & & \multicolumn{3}{c}{Component 2} & & \multicolumn{3}{c}{Component 3} \\
    \cline{2-4} \cline{6-8} \cline{10-12}
    Density & $\pi_1$ & $\mu_1$ & $\sigma_1^2$
            & & $\pi_2$ & $\mu_2$ & $\sigma_2^2$
            & & $\pi_3$ & $\mu_3$ & $\sigma_3^2$ \\
    \cline{2-4} \cline{6-8} \cline{10-12}
    1 & 0.75 & 0.00 & 0.83 & & 0.25 & 1.37 & 0.09 & & \multicolumn{1}{c}{--} & \multicolumn{1}{c}{--} & \multicolumn{1}{c}{--} \\
    2 & 0.45 & -0.93 & 0.22 & & 0.45 & 0.93 & 0.22 & & 0.10 & 0.00 & 0.04 \\
    3 & 0.50 & -0.74 & 0.14 & & 0.30 & 0.37 & 0.55 & & 0.20 & 1.47 & 0.14 \\
    4 & 0.50 & 0.00 & 0.14 & & 0.35 & 1.28 & 0.14 & & 0.15 & 2.56 & 0.11 \\
    \hline
  \end{tabular}
\caption{Top panel: univariate density functions used in the simulation study. Bottom panel: corresponding parameter settings for the mixture components.}
\label{tab:tabSimulationSetting}
\end{minipage}
\end{table}

In Table \ref{tab:results}, we report Monte Carlo estimates of the MSCS coverage probabilities, the corresponding MSCS sizes, and the probability that the BIC order $\hat k$ coincides with the true order $k_0$. These results align with our theoretical expectations, showing that coverage probability approaches or exceeds the nominal level $1-\alpha$ as $n$ increases across all scenarios. However, different densities exhibit distinct behaviors: while Densities 1 and 4 achieve reasonable coverage even at $n = 100$, Densities 2 and 3 are more challenging.

\begin{table}[t]
    \centering
    \caption{Monte Carlo estimates of MSCS coverage probability and size (in parentheses) for significance level $\alpha = \{0.01, 0.05, 0.1\}$, based on $\delta_n(\hat{k}, k) = p_{k}  -  p_{\hat k}$. The row $\hat{k} = k_0$ shows Monte Carlo estimates of $\mathbb{P}(\hat k=k_0)$. Calculations are based on $B = 500$ Monte Carlo samples of size $n$. }
    \begin{tabular}{lcccccc}
        \toprule
        \multirow{2}{*}{} & \multirow{2}{*}{} & & Density 1 & Density 2& Density 3 & Density 4 \\
         $n$ & $\alpha$ &  & $k_0 = 2$ & $k_0 = 3$ &$k_0 = 3$ & $k_0 = 3$ \\
        \midrule
        \multirow{6}{*}{100}  
        & \multirow{2}{*}{0.10}  & Coverage (\%) & 98.0 (4.9) & 66.8 (4.4) & 65.4 (3.8) & 89.6 (5.6) \\
        &  & $\hat{k} = k_0$ (\%) & 55.0 & 3.0 & 2.2 & 24.8 \\
        & \multirow{2}{*}{0.05} & Coverage (\%) & 98.8 (5.9) & 74.2 (5.8) & 66.0 (4.3) & 90.4 (7.6) \\
          &  & $\hat{k} = k_0$ (\%) & 52.4 & 3.4 & 3.2 & 20.2 \\
        & \multirow{2}{*}{0.01} & Coverage (\%) & 99.8 (8.6)  & 82.8 (8.5) & 80.4 (7.0) & 92.2 (9.5) \\
        &  & $\hat{k} = k_0$ (\%) & 53.6 & 2.2 & 3.0 & 26.0 \\
        \midrule
        \multirow{6}{*}{250}  
        & \multirow{2}{*}{0.10}  & Coverage (\%) & 99.8 (2.9) & 74.0 (4.2) & 76.8 (4.3) & 95.8 (3.5) \\
        &  & $\hat{k} = k_0$ (\%) & 94.2 & 4.6 & 5.8 & 73.8 \\
        & \multirow{2}{*}{0.05} & Coverage (\%) & 100 (3.8) & 79.4 (5.8) & 80.2 (5.4) & 95.6 (4.7) \\
        &  & $\hat{k} = k_0$ (\%) & 93.6 & 4.0 & 3.8 & 69.6 \\
        & \multirow{2}{*}{0.01} & Coverage (\%) & 100 (6.3) & 87.4 (8.2) & 89.0 (9.0) & 95.0 (7.8) \\
        &  & $\hat{k} = k_0$ (\%) & 92.4 & 5.0 & 4.2 & 68.0 \\
        \midrule
        \multirow{6}{*}{1000}  
        & \multirow{2}{*}{0.10}  & Coverage (\%) & 99.8 (1.6) & 99.4 (5.5) & 97.4 (5.2) & 96.4 (1.5) \\
        &  & $\hat{k} = k_0$ (\%) & 99.8 & 34.2 & 36.4 & 96.4 \\
        & \multirow{2}{*}{0.05} & Coverage (\%) & 99.8 (2.1) & 99.4 (7.0) & 98.6 (6.5) & 98.8 (1.8) \\
        &  & $\hat{k} = k_0$ (\%) & 99.8 & 28.8 & 36.4 & 98.4 \\
        & \multirow{2}{*}{0.01} & Coverage (\%) & 100 (3.2) & 100 (9.2) & 98.6 (9.2) & 99.2 (2.5) \\
        &  & $\hat{k} = k_0$ (\%) & 99.6 & 32.6 & 33.2 & 99.0 \\
        \bottomrule
    \end{tabular}
    \label{tab:results}
\end{table}

These difficulties are also reflected in the MSCS size. In simpler cases, the MSCS is relatively small and tends to converge to the single true model as $n$ increases. In more complex cases, even with $n = 1000$, the sets widen to ensure inclusion of $k_0$, reflecting the frequent failure of BIC to select the true mixture order. The comparison between the percentage of correct BIC selections ($\hat{k} = k_0$) and the MSCS coverage underscores the advantages of the proposed approach. The MSCS accounts for selection uncertainty, as evidenced by the sizes of the sets -- an aspect entirely overlooked by traditional single-model selection. Moreover, it mitigates BIC selection errors in finite samples. This is particularly evident in complex settings where the BIC performs poorly even for large $n$, highlighting the benefits of a multi-model approach. Overall, our results emphasize the importance of addressing selection uncertainty when choosing the number of components, especially in challenging scenarios where reliance on a single model might be misleading.

\paragraph{Acidity data.} We analyze the Acidity dataset, which contains log-scale acidity index measurements from $155$ lakes in the Northeastern United States \citep{crawford1994application}. From a Bayesian perspective, \citet{richardson1997bayesian} report sizeable posterior probabilities for models with $2$ to $6$ components. However, their sensitivity analysis shows that prior choices can heavily influence the posterior distribution, with some settings supporting models with more than $10$ components. On the frequentist side, \citet[][Sec. 6.6.2]{McLachlan_Mixture} apply a resampling-based likelihood ratio test, suggesting that a $3$-component model should not be exceeded, although they do not address explicitly selection uncertainty. These discrepancies have established this dataset as a benchmark for evaluating mixture-based density estimation.

\begin{figure}[t]
    \centering
    \includegraphics[scale = 0.61]{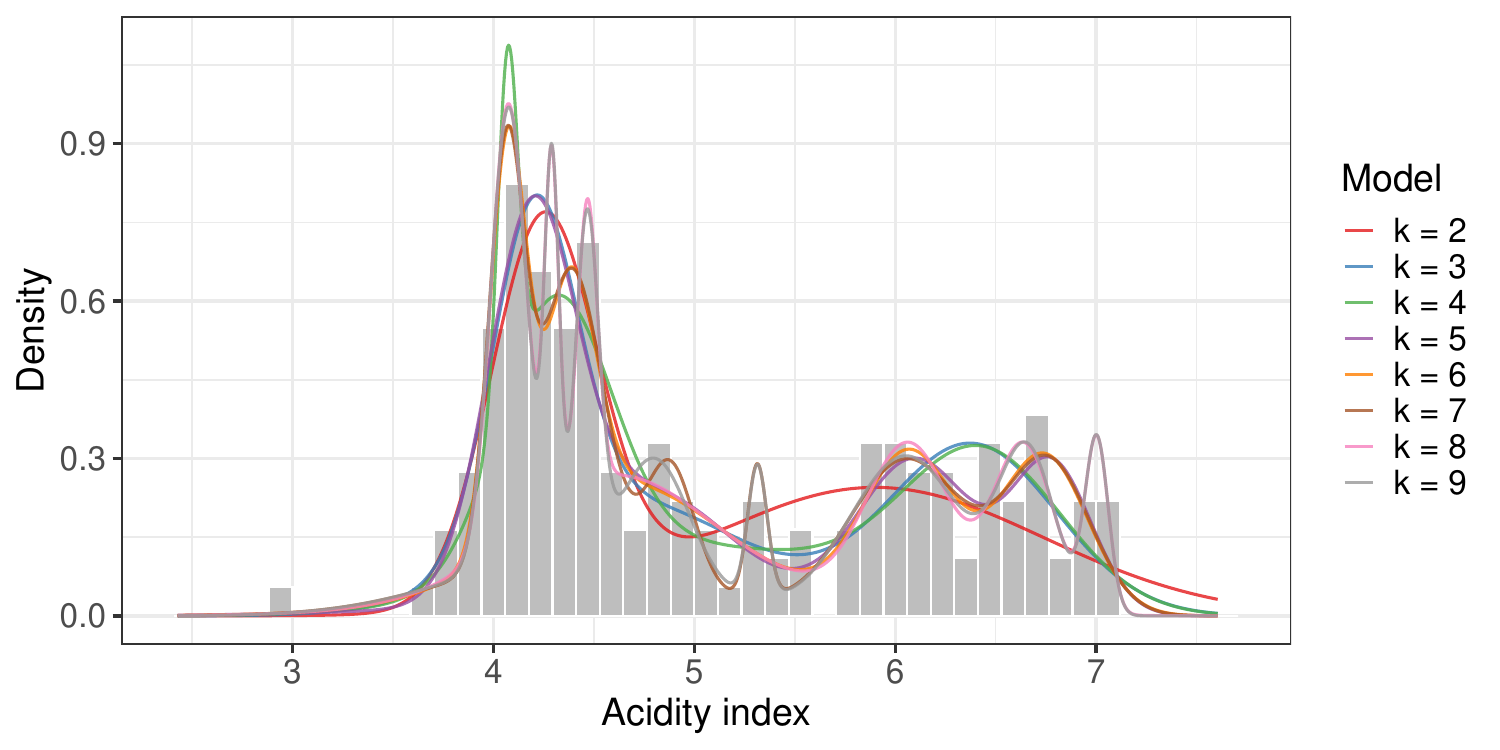}
    \caption{Histogram of the acidity index with overlaid estimated densities from the models included in the 95\% MSCS, constructed using BIC as the reference criterion, and $\delta_n = p_k - p_{\hat k}$.}
    \label{fig:acidityEstimates}
\end{figure}

In our analysis, we construct a $95\%$ MSCS. Consistent with the Monte Carlo experiments, we select $\hat{k}$ using the BIC and apply the correction factor $\delta_n(\hat{k}, k) = p_{\hat k} - p_k$. The BIC model is $\hat{k} = 3$, which is automatically included in the MSCS, and the resulting MSCS is $\hat{\Gamma} = \{2, 3, \dots, 9\}$, with corresponding fitted densities  displayed in Figure \ref{fig:acidityEstimates}. These results highlight substantial model selection uncertainty, which is overlooked when relying solely on the single BIC model or the resampling  approach of \citet{McLachlan_Mixture}.

\section{Discussion and future directions}\label{sec:discussion}

The proposed set estimator provides a principled approach to addressing uncertainty in mixture order selection, and offers a new approach to overcome the limitations of single-model selection. Our numerical studies highlight the risk of model misspecification, even for selection-consistent criteria like BIC in large-sample settings. The MSCS approach mitigates this risk by identifying a set of plausible models, which remains relatively small when BIC is used as the reference. As shown in the proof of Theorem \ref{thm:coverage}, we establish that $
\mathbb{P}(k_0 \in \hat{\Gamma}) \geq 1 - \alpha\mathbb{P}(\hat{k} \neq k_0)$
for sufficiently large $n$. This suggests that the nominal coverage level \( 1 - \alpha \) might be conservative when $\hat{k}$ is selected using a consistent method.  The probability $\mathbb{P}(\hat{k} \neq k_0)$ is expected to influence the MSCS size by affecting the power of the test. Future research should investigate how different selection procedures affect this probability and, in turn, the size and coverage properties of the resulting MSCS. In addition, moving beyond first-order asymptotics via bootstrap calibration would be especially useful in mixture models, where the null distribution of the likelihood ratio statistic is challenging due to non-identifiability and boundary effects.

Finally, another promising direction is the generalization of our MSCS procedure using test statistics based on goodness-of-fit measures beyond KL divergence. These approaches are particularly appealing  for their broader applicability and their robustness to model misspecification and boundary issues. For example, the hypothesis testing framework of \citet{wichitchan2019hypothesis} supports valid inference under weaker regularity conditions and could serve as a foundation for applying the MSCS methodology to more complex or non-regular settings -- such as multivariate Gaussian mixtures or skewed distributions. 

\bibliographystyle{apalike}
\bibliography{biblio}

\end{document}